\documentclass{modified_ws-procs9x6}

\def\beq{\begin{equation}}
\def\eeq{\end{equation}}
\def\lsim{\mathrel{\rlap{\lower3pt\hbox{\hskip0pt$\sim$}}
    \raise1pt\hbox{$<$}}}         
\def\gsim{\mathrel{\rlap{\lower4pt\hbox{\hskip1pt$\sim$}}
    \raise1pt\hbox{$>$}}}         
\def\simlt{\mathrel{\raise.3ex\hbox{$<$\kern-.75em\lower1ex\hbox{$\sim$}}}}
\def\simgt{\mathrel{\raise.3ex\hbox{$>$\kern-.75em\lower1ex\hbox{$\sim$}}}}

\begin{document}

\title{ Higgs Quark Couplings in SUSY with CP and Flavor Violations}

\author{Durmu{\c s}~A. DEMIR\footnote{{ \uppercase{P}resent \uppercase{A}ddress:
\uppercase{D}epartment of \uppercase{P}hysics,
 \uppercase{I}zmir \uppercase{I}nstitute of
\uppercase{T}echnology, \uppercase{T}urkey, \uppercase{TR}35437}}}

\address{William I. Fine Theoretical Physics Institute,\\ University of Minnesota,
MN55455, USA\\
E-mail: demir@physics.iztech.edu.tr}

\maketitle

\abstracts{ In minimal supersymmetric model with a light Higgs
sector, explicit CP violation and most general flavor mixings in
the sfermion sector, integration of the superpartners out of the
spectrum induces potentially large contributions to the Yukawa
couplings of light quarks via those of the heavier ones. These
corrections can be sizeable even for moderate values of
$\tan\beta$, and remain nonvanishing even if all superpartners
decouple. Then Higgs boson couplings to light quarks assume
spectacular enhancements; in particular, couplings to down and
strange quarks become degenerate with that to the bottom quark.
There arise strikingly non-standard effects that can show up in
both Higgs boson searches and FCNC observables.}

The primary goal of the existing and planned colliders and of the
meson factories is to test the standard model (SM) and determine
possible 'new physics' effects on its least understood sectors:
breakdown of CP, flavor and gauge symmetries. In the standard
picture, both CP and flavor violations are restricted to arise
from CKM matrix, and the gauge symmetry breaking is accomplished
by introducing the Higgs field. However, the Higgs sector is badly
behaved at quantum level; its stabilization against quadratic
divergences requires supersymmetry (SUSY) or some other extension
of the standard model (SM). The soft breaking sector of the MSSM
accommodates novel sources for CP and flavor violations
\cite{cpv,fv}. The Yukawa couplings, which are central to Higgs
searches at the LHC, differ from all other couplings in the
lagrangian in one aspect: the radiative corrections from sparticle
loops depend only on the ratio of the soft masses and hence they
do not decouple even if the SUSY breaking scale lies far above the
weak scale. In this sense, non-standard hierarchy and texture of
Higgs-quark couplings, once confirmed experimentally, might
provide direct access to sparticles irrespective of how heavy they
might be. This talk is intended to summarize the results of recent
work \cite{demir} which discusses the radiative corrections to
Yukawa couplings from sparticle loops and their impact on FCNC
observables and Higgs phenomenology.

The soft breaking sector mixes sfermions of different flavor via
the off--diagonal entries of the sfermion mass--squared matrices.
The LR and RL blocks are generated after the electroweak breaking
with the maximal size ${{O}}(m_t M_{SUSY})$, and their flavor
mixing potential is dictated by the Yukawa couplings ${\bf
Y}_{u,d}$ and by the trilinear coupling matrices ${\bf
Y}_{u,d}^{A}$ with $ \left({\bf Y}_{u,d}^{A}\right)_{i j} =
\left({\bf Y}_{u,d}\right)_{i j} \left(A_{u,d}\right)_{i j} $
where $A_{u,d}$ are not necessarily unitary so that even their
diagonal entries contribute to CP--violating observables. The
flavor mixings in LL and RR sectors, however, are insensitive to
electroweak breaking; they are of pure SUSY origin. Clearly, CP
violation in LL and RR sectors is restricted to the
flavor-violating entries due to hermiticity.

The effective theory below the SUSY breaking scale $M_{SUSY}$
consists of a modified Higgs sector; in particular, the tree level
Yukawa couplings receive important corrections from sparticle
loops. For instance, the $d$ quark Yukawa coupling relates to the
physical Yukawas via
\begin{eqnarray}
\label{corryuk}
h_d &=& \frac{g_2\overline{m_d}}{\sqrt{2} M_W}\ \frac{\tan\beta}{1+ \epsilon_\beta}\
\left[ 1 - \frac{\epsilon_\beta}{1+ \epsilon_\beta}\
\left\{ \frac{\overline{m_s}}{\overline{m_d}} \left(\delta^{d}_{12}\right)_{L R} +
\frac{\overline{m_b}}{\overline{m_d}} \left(\delta^{d}_{13}\right)_{L R} \right\} \right]
\end{eqnarray}
where $\overline{m_i}$ are the running quark masses at $M_{SUSY}$, $\epsilon_{\beta} = \epsilon \tan\beta$,
$\epsilon=(\alpha_s/3\pi) e^{-i(\theta_{\mu}+\theta_g)}$, and
\begin{eqnarray}
\left(\delta^{d}_{ij}\right)_{L R}=\frac{1}{6}\ \left(\delta^{d}_{ij}\right)_{R R} \left(\delta^{d}_{ji}\right)_{LL}
\end{eqnarray}
with the SUSY CP--odd phases defined as $\theta_{g}=
\mbox{Arg}[M_{g}]$, $\theta_{\mu}= \mbox{Arg}[\mu]$,
$\theta^{d}_{ij}=\mbox{Arg}[\left(A_d\right)_{ij}$ so on. The mass
insertions $\left(\delta^{d,u}_{ij}\right)_{R R, L L}$ are defined
as the ratio of $(i,j)$--th entry of
$\left(M_{D,U}^{2}\right)_{RR, LL}$ to the mean of the diagonal
entries.  The structure in (\ref{corryuk}) repeats for other quark
flavors as well. In contrast to the minimal flavor violation (MFV)
scheme, the Yukawa couplings acquire large corrections from those
of the heavier ones as suggested by the terms in square bracket.
Indeed, the radiative corrections to $h_d/\overline{h_d}$,
$h_s/\overline{h_s}$, $h_u/\overline{h_u}$ and
$h_c/\overline{h_c}$ involve, respectively, the large factors
$\overline{m_b}/\overline{m_d}\sim (\tan\beta)_{max}^{2}$,
$\overline{m_b}/\overline{m_s}\sim (\tan\beta)_{max}$,
$\overline{m_t}/\overline{m_u} \sim (\tan\beta)_{max}^{3}$, and
$\overline{m_t}/\overline{m_c}\sim (\tan\beta)_{max}^{2}$ with
$(\tan\beta)_{max}\simlt \overline{m_t}/\overline{m_b}$. Unlike
the light quarks, the top and bottom Yukawas remain stuck to their
MFV values to a good approximation. Therefore, the SUSY flavor
violation sources mainly influence the light sector whereby
modifying several processes they participate. These corrections
are important even at low $\tan\beta$. As an example, consider
$\left(\delta^{d}_{13}\right)_{L R}\sim 10^{-2}$ for which
$h_d/h_d^{MFV} \simeq 0.02 (2.11), -2.3 (6.6), -4.6 (17.7)$ for
$\tan\beta=5, 20, 40$ at $\theta_{\mu}+\theta_g\leadsto 0 (\pi)$.
Note that the Yukawas are enhanced especially for
$\theta_{\mu}+\theta_g\leadsto \pi$ which is the point preferred
by Yukawa--unified models such as SO(10). In general, as
$\tan\beta\rightarrow (\tan\beta)_{max}$ the Yukawa couplings of
down and strange quarks become amproximately degenerate with the
bottom Yukawa for $\left(\delta^{d}_{13, 23}\right)_{L R}\sim 0.1$
and $\theta_{\mu}+\theta_g\leadsto \pi$. There is no $\tan\beta$
enhancement for up quark sector but still the large ratio
$\overline{m_t}/\overline{m_u}$ sizeably folds $h_u$ compared to
its SM value: $h_u \simeq 0.6\ e^{i(\theta^{u}_{11} - \theta_g)}\
\overline{h_c}$ with $\left(\delta^{u}_{13} \right)_{L R}\sim
0.1$.

The SUSY flavor violation influences the Higgs-quark interactions by ($i$) modifying
$H^a \overline{q} q$ couplings via sizeable changes in Yukawa couplings, and by ($ii$)
inducing large flavor changing couplings $H^a \overline{q} q'$:
\begin{eqnarray}
\label{higgsquark}
&& \frac{\overline{h_{d^i}}^{SM}}{\sqrt{2}}
\left[ \frac{h_d^i}{\overline{h_d^i}}\
\tan\beta\ C^{d}_a + \left(\frac{h_d^i}{\overline{h_d^i}} -1 \right) \left(e^{i(\theta^{d}_{ii}+\theta_{\mu})}\, C^{d}_{a} -
C^{u \star}_a \right)\right]\ \overline{d^i_R}\ d^i_L\ H_a \nonumber\\
&+& \frac{\overline{h_{d^i}}^{SM}}{3 \sqrt{2}} \epsilon \tan\beta
\left[ \frac{h_d^i}{\overline{h_d^i}}\ \left(\delta^{d}_{ij}\right)_{L L} +
\frac{h_d^j}{\overline{h_d^i}}\ \left(\delta^{d}_{ij}\right)_{R R} \right]
\left( \tan\beta\ C^{d}_a - C^{u \star}_a\right)
\overline{d^i_R}\ d^j_L\ H_a\nonumber
\end{eqnarray}
where $C^{d}_{a}\equiv \{ - \sin\alpha,$ $\cos\alpha, i\sin\beta,$
$-i\cos\beta\}$ and $C^{u}_{a}\equiv \{ \cos\alpha,$ $\sin\alpha,$
$i\cos\beta, i\sin\beta\}$ in the basis $H_{a}\equiv \{h,H,A,G\}$
if the CP violation effects in the Higgs sector are neglected
(which can be quite sizeable \cite{higgscpv} and add additional
CP-odd phases \cite{add} to Higgs-quark interactions). Similar
structures also hold for the up sector. The interactions contained
in (\ref{higgsquark}) have important implications for both FCNC
transitions and Higgs decay modes. The FCNC processes are
contributed by both the sparticle loops ($e.g.$ the gluino-squark
box diagram for $K^0$--$\overline{K^0}$ mixing) and Higgs exchange
amplitudes. The constraints on various mass insertions can be
satisfied by a partial cancellation between these two
contributions if $M_{SUSY}$ is close to the weak scale. (This
parameter domain needs a global analysis of FCNC constraints to
determine for what ranges of SUSY parameters the bounds on mass
insertions are relaxed.) On the other hand, if $M_{SUSY}$ is high
then the only surviving SUSY contribution is the Higgs exchange.
The main question to be answered is: Is it possible to saturate
FCNC bounds with ${{O}}(1)$ mass insertions? Concerning this
point, consider $B_d\rightarrow \mu^+\mu^-$ decay, which has a
rather small SM background, for $M_{SUSY}\gg m_t$. Using the
explicit expressions of Yukawa couplings (\ref{corryuk}) in
(\ref{higgsquark}) one finds that the Higgs exchange contribution
to this decay gets totally suppressed with ${{O}}(1)$ mass
insertions for $\tan\beta\simeq (\tan\beta)_{max}$ and $\phi_{\mu}
+ \phi_{g} \leadsto \pi$. Therefore, in this parameter domain,
though the flavor-changing Higgs decay channels are sealed up the
decays into similar quarks are highly enhanced. For instance,
$\Gamma(h\rightarrow \overline{d} d)/\Gamma(h\rightarrow
\overline{b} b)\simeq
\left(\mbox{Re}\left[{h_d}/{{h_b}}\right]\right)^{2}$ which is
${{O}}(1)$ when $h_d\sim h_b$ as is the case with SUSY flavor
violation. Such enhancements in light quark Yukawas induce
significant reductions in $\overline{b}b$ branching fraction ---
which is a very important signal for hadron colliders to determine
the non-SM structure of the Higgs boson ($h\rightarrow
\overline{b}b$ has $\sim 90\%$ branching fraction in the SM). If
FCNC constraints are saturated without a strong suppression of the
flavor-changing Higgs couplings (which requires $M_{SUSY}$ to be
close to the weak scale) then Higgs decays into dissimilar quarks
get significantly enhanced. For instance, $h\rightarrow
\overline{b}s +\overline{s}b$ can be comparable to $h \rightarrow
\overline{b}b$. (See \cite{herrero} for a diagrammatic analysis of
$\rightarrow \overline{b}s +\overline{s}b$ decay.) In conclusion,
as fully detailed in \cite{demir}, SUSY flavor and CP violation
sources significantly modify Higgs--quark interactions whereby
inducing potentially large effects that can be discovered at
hadron colliders as well as  meson factories.

\vspace{0.5cm}
The research was supported by DOE grant DE-FG02-94ER40823 at Minnesota.


\begin{thebibliography}{0}
\bibitem{cpv}
M. Dugan, B. Grinstein and L. Hall,
{\it Nucl.\ Phys.\ B} {\bf 255} (1985) 413;
D. Demir, A. Masiero and O. Vives,
{\it Phys.\ Rev.\ D} {\bf 61} (2000) 075009
[arXiv:hep-ph/9909325];
{\it Phys.\ Lett.\ B} {\bf 479} (2000) 230
[arXiv:hep-ph/9911337];
D. Demir,
{\it Nucl.\ Phys.\ Proc.\ Suppl.}  {\bf 101} (2001) 431.

\bibitem{fv}
S. Bertolini $et.al.$,
{\it Nucl.\ Phys.\ B} {\bf 353} (1991) 591;
J. Hagelin, S. Kelley and T. Tanaka,
{\it Nucl.\ Phys.\ B} {\bf 415} (1994) 293.

\bibitem{demir}
D. Demir,
{\it Phys.\ Lett.\ B} {\bf 571}, 193 (2003)
[arXiv:hep-ph/0303249].

\bibitem{higgscpv}
A. Pilaftsis,
Phys.\ Lett.\ B {\bf 435}, 88 (1998)
[arXiv:hep-ph/9805373];
D. Demir,
Phys.\ Rev.\ D {\bf 60}, 055006 (1999)
[arXiv:hep-ph/9901389];
A. Pilaftsis and C. Wagner,
{\it Nucl.\ Phys.\ B} {\bf 553}, 3 (1999)
[arXiv:hep-ph/9902371].

\bibitem{add}
D. Demir,
{\it Phys.\ Rev.\ D} {\bf 60}, 095007 (1999)
[arXiv:hep-ph/9905571].


\bibitem{herrero}
A. Curiel, M. Herrero and D.Temes,
{\it Phys.\ Rev.\ D} {\bf 67}, 075008 (2003)
[arXiv:hep-ph/0210335].


\end{thebibliography}
\end{document}